\documentclass[aps,prb,twocolumn,showpacs,superscriptaddress,groupedaddress,floatfix,preprintnumbers]{revtex4-1}
\usepackage{times}
\usepackage{wrapfig}
\usepackage[pdftex]{graphicx} 
\usepackage{dcolumn}   
\usepackage{bm}        
\usepackage{amssymb}   
\usepackage{amsmath}   
\usepackage{appendix}
\usepackage{color}
\usepackage{parskip}
\usepackage{bbding}
\usepackage{natbib}
\PassOptionsToPackage{caption=false}{subfig}
\usepackage{subfig}

\newcommand{\beq}{\begin{equation}}
\newcommand{\eeq}{\end{equation}}

\newcommand{\beqa}{\begin{eqnarray}}
\newcommand{\eeqa}{\end{eqnarray}}

\newcommand \nline {\nonumber \\}

\newcommand{\mbfr}{{\mathbf r}}

\newcommand \dxdy[2] {\frac{d #1}{d #2}}

\begin{document}

\title{Atomic-scale pathway of early-stage precipitation in Al-Mg-Si alloys}

\author{$^1$Vahid Fallah* (vfallah@uwaterloo.ca),\\ $^2$Andreas Korinek (korinek@mcmaster.ca),\\ $^{3,4}$Nana Ofori-Opoku (nana.ofori-opoku@mail.mcgill.ca),\\
$^5$Babak Raeisinia (Babak.Raeisinia@novelis.com),\\ $^6$Mark Gallerneault (mark.gallerneault@alcereco.com),\\ $^3$Nikolas Provatas (provatas@physics.mcgill.ca),\\ $^1$Shahrzad Esmaeili (shahrzad@uwaterloo.ca)}

\affiliation{$^1$Mechanical and Mechatronics Engineering Department, University of Waterloo, 200 University Avenue West, Waterloo, ON N2L-3G1, Canada}
\affiliation{$^2$Department of Materials Science and Engineering, McMaster University, 1280 Main Street West, Hamilton, ON L8S-4L7, Canada}
\affiliation{$^3$Department of Physics, and Centre for the Physics of Materials, McGill University, 3600 University Street, Montreal, QC H3A-2T8, Canada}
\affiliation{$^4$Department of Mining and Materials Engineering, McGill University, 3610 University Street, Montreal, QC H3A-0C5, Canada}
\affiliation{$^5$Novelis Global Research $\&$ Technology Center, 1950 Vaughn Road, Kennesaw, GA 30144, USA}
\affiliation{$^6$Formerly Novelis Research $\&$ Technology Center, 945 Princess Street, Kingston, ON K7L-5L9, Canada}

\begin{abstract}
Strengthening in age-hardenable alloys is mainly achieved through nano-scale precipitates whose formation paths from the atomic-scale, solute-enriched entities are rarely analyzed and understood in a directly-verifiable way. Here, we discover a pathway for the earliest-stage precipitation in Al-Mg-Si alloys: solute clustering leading to three successive variants of FCC clusters, followed by the formation of non-FCC $GP$-$zones$. The clusters, which originally assume a spherical morphology ($C1$), evolve into elongated clusters and orient themselves on $\{111\}_{Al}$ ($C2$) and subsequently on $\{100\}_{Al}$ planes and $<$$100$$>_{Al}$ directions ($C3$). We also analyze the association of quenched-in dislocations with clustering phenomena. The results of this work can open a new frontier in advancing alloy-process-property design for commercially-important age-hardenable Al alloys.
\end{abstract}

\maketitle


\section{Introduction}
Al-Mg-Si alloys, i.e. AA6000 series, are lightweight medium-strength heat treatable Al alloys with a desirable combination of cost-effective engineering properties that has made them suitable for a wide range of applications from various transportation~\cite{lloyd13} and building industries~\cite{zhang10} to high voltage power transmission~\cite{karabay06}. The ever increasing interest in these alloys, particularly for automotive panel applications, has urged intense investigations on the formation kinetics and evolution of precipitates in such systems~\cite{vasilyev11,wang03,esmaeili03-2,esmaeili04,esmaeili05-2,buha08-2,buha07,pogatscher11,pogatscher12}. Automotive paint bake cycling processes involve early- to medium-stage underaging conditions~\cite{lloyd98,murayama99,esmaeili03-2,pogatscher12}. Recent results on such early- and medium-stage precipitation hardening in Al-Mg-Si alloys have revealed that structure-hardening relationships are strongly dependent on the alloy composition and details of the aging process~\cite{buha07,buha08-2,pogatscher11,pogatscher12}. These factors have been reported to control the nucleation and growth conditions of the hardening phase, $\beta''$, which mainly forms during medium stage of aging~\cite{buha07,buha08-2,pogatscher12,pogatscher13} and is responsible for the highest hardening achieved in the peak-aged condition, while also contributing to hardening in underaged conditions~\cite{zandbergen97,murayama99,andersen98,edwards98,marioara03,esmaeili03-3,esmaeili03,esmaeili05}.    

Dedicated studies have attempted to analyze the early-stage precipitation in 6000 series Al alloys in terms of the sequential nature and compositional evolution of the so-called solute clusters and GP-zones. However, analyses of such phenomena have been mainly conducted through indirect measurements such as Differential Scanning Calorimetry (DSC)~\cite{edwards98,gupta99,serizawa08,banhart11,chang11} and Positron Annihilation Spectroscopy (PAS)~\cite{banhart11,lay12}. Semi-direct analyses by atom probe methodologies have also been used for several investigations~\cite{murayama99,edwards98,buha07,esmaeili07,serizawa08,buha08-2,marceau13}. These investigations have led to the identification of multiple clustering processes which start to form immediately after quenching~\cite{edwards98,gupta99,murayama99,serizawa08,banhart11}. Although variations in the compositional characteristics of these clusters have been reported in the past, most recent studies indicate that the first group of clusters are Si-rich aggregates~\cite{buha08-2,banhart11}. The second group of clusters, speculated to be descendants of the first Si-rich clusters, is shown to exhibit Mg-enrichment while growing~\cite{serizawa08,buha08-2,banhart11,martinsen12}. It has been reported that a larger Si content in the alloy composition promotes cluster nucleation and leads to a finer distribution of $\beta''$ precipitates at later stages of aging~\cite{ohmori02,marioara05,buha08-2}. The morphology of early-stage precipitates is generally reported from atom probe investigations to be spherical~\cite{esmaeili07,serizawa08,buha08-2,marceau13}.
A few studies have also reported GP-zones with spherical morphology and unknown crystal structure~\cite{dutta91,murayama99,buha07}. Dedicated studies have been conducted to analyze the role of quenched-in vacancies in early stage precipitation phenomena in these alloys and a number of related hypotheses have been proposed~\cite{esmaeili04,esmaeili07,banhart11,chang11,vasilyev11,lay12}.     

Despite the vast amount of published information, there continues to remain important unanswered questions about the structural/morphological evolution of early clusters as well as the fundamental mechanisms and characteristics of diffusional transformations that occur at the earliest stages of precipitation in Al-Mg-Si alloys. Although indirect methods such as electrical resistivity and semi-direct atom probe studies have shown evidence for cluster formation, these clusters have not been directly imaged in a systematic way and no sequence of evolution in clustering itself has been reported. All previous works have suggested that, up until the formation of needle-like pre-$\beta''$ $GP$-$zones$~\cite{marioara01}, there exist only spherical clusters with an FCC structure similar to the matrix, followed by spherical $GP$-$zones$~\cite{dutta91,murayama99,buha07} which have been poorly shown and never been structurally or chemically analyzed. Furthermore, there exists a significant gap of knowledge in theoretical analysis of solute clustering phenomena, and the roles that quenched-in dislocations may play during early-stage transformations in these alloys. The gap also includes directly-verifiable (i.e. large-scale) atomistic simulation strategies that are essential for the development of future break-through strategies in alloy and process designs. The aim of this work is to move a major step forward in filling an important gap in this knowledge through both atomic-scale observations and computational modelling. The computational approach is based on the recently-developed modeling paradigm, known as the Phase Field Crystal ({\it PFC}) methodology~\cite{elder04,elder07,greenwood10,greenwood11,greenwood11-2}, which has shown a remarkable capacity in capturing the salient physics of diffusional phase transitions involving atomic scale elastic and plastic interactions~\cite{fallah13-2,fallah13,ofori13,fallah12}. Using the {\it PFC} approach, Fallah {\it et al.}~\cite{fallah12,fallah13}, for the first time, analyzed atomistic transformations and the full free energy-based path leading to the earliest stage precipitation in the presence of quenched-in and mobile dislocations in the Al-Cu-(Mg) alloy system. They also used high resolution transmission and scanning-transmission electron microscopy (HRTEM and HRSTEM, respectively) methods to assess the 3D atomic-scale simulation results through direct observation of clustering that occurs during natural aging in the Al-Cu system~\cite{fallah13-2}. That study well demonstrated the validity of the simulated dislocation-induced early-stage clustering phenomena and the preferred orientation of the early clusters on the FCC closed packed (i.e. $\{111\}_{Al}$) planes. However, the possibility for other orientations was not analyzed. This work reports a pioneering systematic study based on atomic-scale direct observations and {\it PFC} approach to analyze clustering phenomena in Al-Mg-Si alloy system.

In this study, early-stage precipitation phenomenon is studied for solutionized/aged Al-Mg-Si alloys using HRTEM and scanning transmission electron microscopy (STEM), along with atomic-scale 3D {\it PFC} simulations conducted on a continuous isothermal aging basis. We not only report direct imaging of the earliest clusters, but also we identify, for the first time, that clusters themselves undergo a morphology and orientation evolution prior to evolving to non-FCC $GP$-$zones$. In other words, we introduce the clustering missing links in the evolutionary path of early-stage precipitation, connecting the early spherical clusters with FCC structure to non-FCC $GP$-$zones$, which may then act as precursors to the previously reported pre-$\beta''$ $GP$-$zones$ with a well-defined C-centred monoclinic structure~\cite{marioara01}. The experimental and numerical analysis of early-stage precipitation in this study, however, is focused only on the evolution of clusters with an FCC structure similar to the matrix. In reporting the analysis, our convention is to name all early precipitates with a lattice structure similar to the matrix lattice (i.e. FCC) as \textquotedblleft $clusters$\textquotedblright, and those which deviate from the FCC structure as \textquotedblleft $GP$-$zones$\textquotedblright. 

\section{Experimental}
High purity Al-0.8Mg-0.8Si and Al-0.94Mg-0.47Si ($at.\%$) alloys (called $A1$ and $A2$ alloys, respectively, hereafter), representing two widely different atomic Mg/Si ratios of 1 and 2, were supplied by Novelis Inc. as 1 mm thick cold-rolled sheets. They were solution treated in an air furnace at 560~$^\circ C$ for 20 min, quenched in water and immediately aged at 180~$^\circ C$ for various times, i.e. 5, 15 and 20 min. Immediately after the heat treatment, the samples were mechanically ground, electropolished and kept in liquid nitrogen (LN2) prior to electron microscopy. Electropolished TEM samples were imaged at high resolution using an FEI Titan low base microscope, operated at 300 kV and equipped with a CEOS image corrector.  

\section{Model structure}  
A 3D ternary PFC model was adopted from the multicomponent formalism of structural PFC (XPFC) methodology developed by Ofori-Opoku {\it et al.}~\cite{ofori13}. As also documented in Refs.~\cite{fallah12,fallah13,fallah13-2}, the mean-field approximations of PFC free energy was carried out following the procedure detailed in ref.~\cite{greenwood11-2}. The resultant free energy of the alloy was used to construct the relevant phase diagram and also to estimate the work of formation of clusters through calculation of individual contributing terms (i.e., the driving force, strain energy and surface energy). The descriptions of such methodologies can be found in detail in Refs.~\cite{fallah12,fallah13,fallah13-2}. Below, we present the details of the above {\it PFC} calculations and simulation methodologies for clustering in a ternary alloy system in 3D.    

\subsection{XPFC formalism for a ternary system}
In a three-component system, the free energy functional can be described by the sum of two terms: ideal and excess free energy, each as a function of three individual density fields (i.e., $\rho_{A}$, $\rho_{B}$ and $\rho_{C}$). The ideal term, which represents the local free energy, drives the density fields to a uniform state. On the contrary, the excess term favors periodic density fields, which arises due to species interactions (i.e. between various density fields). The following free energy functional can be written from classical density functional theory (CDFT)~\cite{ofori13}: 
\begin{align}
\frac{\Delta {\mathcal F}}{k_B T \rho^o} \equiv \int d\mbfr~{f}  = \int d\mbfr~\{\Delta F_{id}+\Delta F_{ex} \},
\label{truncated-DFT-energy}
\end{align} 
where $\Delta F_{id}$ and $\Delta F_{ex}$ are, respectively, the dimensionless ideal energy and excess energy, $k_B$ the Boltzmann constant, $T$ the temperature and $\rho^o$ a reference system density (defined below). The ideal free energy of the ternary alloy is given by~\cite{fallah13} 
\begin{align}
\Delta F_{id} &= \rho_{A} \ln \left(\frac{\rho_{A}}{\rho_{A}^{o}}\right)-\delta\rho_{A} + \rho_{B} \ln \left(\frac{\rho_{B}}{\rho_{B}^{o}}\right) - \delta\rho_{B}\nline&+\rho_{C} \ln \left(\frac{\rho_{C}}{\rho_{C}^{o}}\right) - \delta\rho_{C}, 
\label{idealenergy-log}
\end{align}  
where $\rho_i$ (with $i=A,B,C$) is the density of component $i$, $\delta \rho=\rho_i-\rho_i^o$ and ${\rho_{i}^{o}}$ the reference density of component $i$, taken to be that of liquid at solid-liquid coexistence. 

The excess free energy term, described by two-point correlations between atoms (i.e. density peaks), introduces elasticity, crystalline symmetry and gives rise to interactions between topological defects within the solid phases. Following Ofori-Opoku {\it et al}~\cite{ofori13}, this term can be written in the following general form
\begin{align}
\Delta F_{ex} &= -\frac{1}{2}\sum_i\sum_j \Delta F_{ij}\nonumber \\
&=-\frac{1}{2}\int d\mbfr^{\prime}\,\sum_i\sum_j\delta\rho_i\left(\mbfr\right)\,C_2^{ij}\left(\mbfr,\mbfr^\prime\right)\,
\delta\rho_j\left(\mbfr^\prime\right),
\label{two-particle-interaction}
\end{align}
where $C_2^{ij}$ denotes all possible combinations of the two particle correlations between components $i$ and $j$ with $i,j=A,B,C$ (i.e. the sums run over only the three species). 

Following the methodology described in previous alloy PFC models~\cite{elder07,greenwood11}, a total mass density is defined as $\rho = \sum_{i} \rho_{i}$ while the total reference mass density is taken as $\rho^o = \sum_{i} \rho_{i}^o$. Following Provatas and Majaniemi~\cite{provatas10} and Greenwood {\it et al.}~\cite{greenwood11}, $c_{i} = \rho_{i}/\rho$ defines the concentration of each component $i$ and $c_{i}^o = \rho_{i}^o/\rho^o$ represents the corresponding reference compositions. Also, a dimensionless mass density of the form $n = \rho/\rho^o-1$ is defined for convenience. To conserve mass, we have $\sum_i c_i \equiv 1$ (or $c_C=1-c_A-c_B$ and $c_C^o=1-c_A^o-c_B^o$). With these definitions and also following the approximations described in Refs.~\cite{ofori13,fallah13}, the free energy functional can be re-written in the following reduced form for a three-component system in terms of $n$ and $\{c_i\}$ (i.e. for species $A$ and $B$):
\begin{align}
\check{{\cal F}} &= \int d\mbfr~\Bigg\{\frac{n^2}{2} -\eta \frac{n^3}{6} + \chi \frac{n^4}{12} + \omega \, \Delta F_{\text {mix}} \, (n+1)
\nline&-\frac{1}{2}n\int d\mbfr^\prime C_{eff}(|\mbfr-\mbfr^\prime|)\,n^\prime + \frac{\alpha_{A}}{2}|\nabla c_{A}|^2 + \frac{\alpha_{B}}{2}|\nabla c_{B}|^2 \Bigg\},
\label{simplified-Energy}
\end{align}
where $\Delta F_{\text {mix}}$ denotes the ideal entropy of mixing,
\begin{align}
\Delta F_{\text {mix}} &= c_{A}\ln{\frac{c_{A}}{c_{A}^{o}}} 
+ c_{B}\ln{\frac{c_{B}}{c_{B}^o}}\nline &+(1-c_{A}-c_{B})\ln{\frac{(1-c_{A}-c_{B})}{1-c_{A}^o-c_{B}^o}}.
\label{entropy-mixing}
\end{align}

The parameters $\eta$ and $\chi$ in Eq.~\ref{simplified-Energy} are  introduced to control the variation of the ideal free energy away from the reference density $\rho^o$. $\omega$ is introduced to set the variation of the entropy of mixing away from the reference compositions $c_{A}^o$ and $c_{B}^o$. The coefficients $\alpha_{A}$ and $\alpha_{B}$ set the scale and energy of compositional interfaces. Here, we treat these parameters as free coefficients to quantitatively shape the free energy functional to reproduce desired materials properties, i.e. the equilibrium phase diagram of a given alloy system. 

The correlation function, $C_{eff}$, includes contributions from cross correlation functions of the form 
$C_{ij}$ in the excess energy. Extending the binary formalism of Ref.~\cite{greenwood11} to the case of ternary alloys, we define the following effective correlation function
\begin{align}
C_{eff} &=  X_1 C_2^{AA} +  X_2 C_2^{BB} +  X_3 C_2^{CC},
\label{CorrEff}
\end{align}
where the coefficients $X_i$ are polynomial functions, which interpolate between two-point correlation functions of the pure species through weighing each by the local compositions. In this study, they are defined by
\begin{align}
X_1 &= 1-3c_{B}^2+2c_{B}^3-3(1-c_{A}-c_{B})^2+2(1-c_{A}-c_{B})^3\nline &-4c_{A}c_{B}(1-c_{A}-c_{B})-X_{\lambda},\nonumber \\[1.5ex] 
X_2 &= 1-3c_{A}^2+2c_{A}^3-3(1-c_{A}-c_{B})^2+2(1-c_{A}-c_{B})^3\nline &-4c_{A}c_{B}(1-c_{A}-c_{B})-X_{\lambda},\nonumber\\[0.5ex]
X_3 &= 1-3c_{A}^2+2c_{A}^3-3c_{B}^2+2c_{B}^3-4c_{A}c_{B}(1-c_{A}-c_{B})\nline &-X_{\lambda},  
\label{interp-func}
\end{align}
where $X_{\lambda}=\lambda c_Ac_B+\lambda c_A(1-c_A-c_B)+\lambda c_B(1-c_A-c_B)$. With these interpolation functions, the variation of free energy of the solid phases is determined by structural changes due to compositional variations. At $\lambda=0$ (equivalent to $X_{\lambda}=0$), Eq.~\ref{interp-func} satisfies $X_1+X_2+X_3 \equiv 1$ at all compositions~\cite{fallah13}. At positive $\lambda$ values, the term $X_{\lambda}$ allows an additional degree of freedom for robust fitting of the free energy landscapes of the solid phases when attempting to reproduce the system equilibrium properties (i.e. the phase diagram reconstruction).

The two-point correlation functions in Fourier space, $\hat{C}^{ii}_2(\vec{k})$, are set by peaks at $k_{j}$ corresponding to the inverse of interplanar spacings of the main reflection from the $j^{\rm th}$ family of planes in the unit cell of the crystal structure favoured by component $i$. In reciprocal space, each peak  is represented by a Gaussian form of width $\alpha_j$ described below.  
\beq
\hat{C}^{ii}_{2j}=e^{-\frac{\sigma^2}{\sigma^2_{Mj}}}e^{-\frac{(k-k_j)^2}{2\alpha^2_j}}. 
\label{CorrF}
\eeq

The height of the peak is modulated by a Debye-Waller-like prefactor, which is set by an effective temperature $\sigma$ and a transition temperature $\sigma_{Mj}$~\cite{greenwood11-2}. The effective temperature $\sigma$, or the model's reduced temperature, is a dimensionless variable which emulate the effect of temperature in our modelling paradigm. Accordingly, the effective transition temperature, $\sigma_{Mj}$, is a reference melting temperature which can be set for each individual phase $j$ in the system when reconstructing the experimental phase diagram. 

The dynamical equations of motion for density and concentration fields follow dissipative dynamics~\cite{fallah13}:
\begin{align}
\frac{\partial n}{\partial t} &= \nabla \cdot M_n\nabla \frac{\delta \check{{\cal F}}}{\delta\,n} \nonumber \\
&=\nabla\cdot M_n \nabla\Biggl\{n-\eta \frac{n^2}{2} + \chi \frac{n^3}{3} + \omega\Delta F_{\text {mix}}-C_{eff}\,n^\prime\Biggr\},
\label{dynamics-n}
\end{align}
\begin{align}
&\frac{\partial c_A}{\partial t} = \nabla \cdot M_{{c}_{A}}\nabla \frac{\delta \check{{\cal F}}}{\delta c_A}\nonumber \\
&=\nabla\cdot M_{{c}_{A}}\nabla\Biggl\{\omega(n+1)\frac{\partial \Delta F_{\text {mix}}}{\partial c_A}
-\frac{1}{2}n\frac{\partial C_{eff}}{\partial c_A}\,n^\prime - \alpha_A\nabla^2c_A \Biggr\},\nonumber \\
\label{dynamics-cA}
\end{align}
\begin{align}
&\frac{\partial c_B}{\partial t} = \nabla \cdot M_{{c}_{B}}\nabla \frac{\delta \check{{\cal F}}}{\delta c_B}\nonumber \\
&=\nabla\cdot M_{{c}_{B}}\nabla\Biggl\{\omega(n+1)\frac{\partial \Delta F_{\text {mix}}}{\partial c_B}
-\frac{1}{2}n\frac{\partial C_{eff}}{\partial c_B}\,n^\prime - \alpha_B\nabla^2c_B \Biggr\}.\nonumber \\
\label{dynamics-cB}
\end{align}

$M_n$, $M_{c_{A}}$ and $M_{c_{B}}$ are dimensionless mobility coefficients for density and compositions fields. In principle, these coefficients themselves must be considered as functions of density, concentration and temperature. In our isothermal simulations, however, the mobility coefficients are estimated as constants (i.e. set to 0.01 for density and 10 for each concentration field, to introduce a faster diffusion of solute atoms compared to the mobility of crystal defects such as dislocations).

\subsection{Mean-field approximation of free energy} 
For the purpose of phase diagram reconstruction and classical energy analysis of  nucleation of clusters (i.e. by calculating the individual contributing terms such as driving force, surface energy and strain energy), we approximate the free energy functional in Eq.\ref{simplified-Energy} as a function of the two concentration fields and temperature. This is done within the framework of the mean-field {\it PFC} approximations~\cite{greenwood11-2}. Here, we calculate the free energy of the solid phase by introducing the following two-mode approximation of the corresponding density field into Eq.\ref{simplified-Energy} and integrating over the unit cell volume~\cite{greenwood11-2}:
\begin{align}
n_i(\vec{r})=\sum_{j=1}^{N_i}A_j\sum_{l=1}^{N_j}e^{2\pi\mathbf{i}\vec{k}_{l,j}.\vec{r}/a_i},
\label{Density}
\end{align}
where the subscript $i$ denotes a particular solid phase with a lattice spacing $a_i$, and the index $j$ counts the number of modes $1 \cdots N_i$ in the $i$-phase. $A_j$ is the amplitude of mode $j$ and the index $l$ counts the $N_j$ reciprocal space peaks representing mode $j$. $\vec{k}_{l,j}$ is then the reciprocal lattice vector corresponding to the index $l$ in family $j$, normalized to a lattice spacing of 1. The lattice spacing, $a_i$, is taken as a variable weighted by the two concentration fields (i.e. $c_{Mg}$ and $c_{Si}$) using the interpolation functions defined in Eqs.~\ref{interp-func}. This way, there will be a rise in the energy of solid for the concentrations away from that of the pure species $i$. In case of a similar crystal structure for all species (i.e. differing only in the lattice parameter), such a variable lattice parameter introduces energy barriers as the concentration changes from one species towards the others. The resulting free energy of the crystalline phase is then minimized for the (non-conserved) amplitudes $A_j$. The minimization methodology is described in more detail in Refs.~\cite{greenwood11-2,ofori13}.

\begin{figure*}[htbp]
\resizebox{6.2in}{!}{\includegraphics{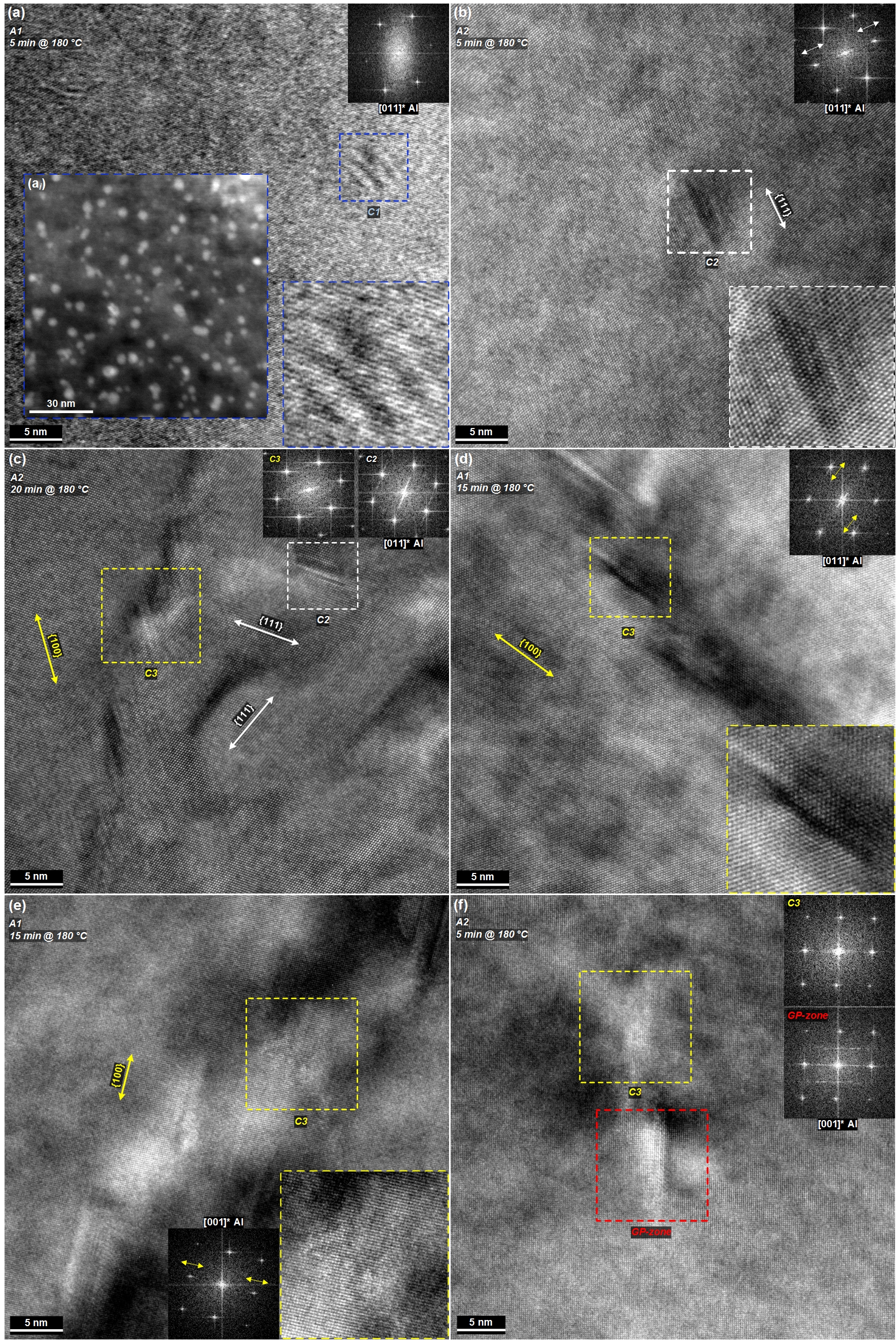}}
\caption{\textbf{HRTEM micrographs showing the evolution of clusters/precipitates.} (a) $C1$, (b) $C2$, (c) $C2$-$C3$ coexistence, (d-e) $C3$ and (f) $C3$-$GP$-$zone$ coexistence; ($a_i$) shows the STEM image of (a) in $[011]_{Al}$ zone axis; The inset FFTs are those of the boxed regions.}
\label{fig:HRTEMcl}
\end{figure*}

\section{Evolution of early clusters/precipitates}
Here, we present both experimental and simulated evolution of early clusters/precipitates in Al-Mg-Si alloys. The experimental observations of the atomic structure of early clusters provide crystallographic information as input to the {\ PFC} simulation of clustering. The simulated evolution of early clusters will be examined and validated by the experimentally observed evolution of clusters throughout the early-stage clustering process, i.e. when alloys aged at 180$^\circ C$ for various time up to 20 min.

\subsection{Characterization of early clusters/precipitates}
HRTEM and STEM analyses identify three main types of clusters, $C1$, $C2$ and $C3$, as illustrated in the selected images shown in Fig.~\ref{fig:HRTEMcl}. The smallest cluster, $C1$, appears mainly at the earliest stage of aging (i.e. after 5 min aging at $180~^\circ C$) with a spherical morphology, as can be seen in Fig.~\ref{fig:HRTEMcl}a and its inset a$_i$. $C2$ is a larger cluster lying on the closed packed $\{111\}_{Al}$ planes (Fig.~\ref{fig:HRTEMcl}b-c). The largest cluster type, $C3$, observed mainly at a later stage of aging (i.e. after 15 min aging at 180$~^\circ C$), lies on the $\{100\}_{Al}$ planes and is elongated along the $<$$100$$>_{Al}$ directions (Fig.~\ref{fig:HRTEMcl}c-f). The inserted FFT patterns of all $C2$ and $C3$ clusters revealed streaking on the main $FCC_{Al}$ reflections in a direction perpendicular to their elongated orientations. This is indicative of an FCC structure of the clusters whose lattice parameter slightly deviates from that of the Al-rich matrix. The difference in the lattice parameter is in turn linked to a slight compositional deviation from the average matrix composition. The detailed characteristics of the above clusters are summarized in Table~\ref{table:clusters}. As Table~\ref{table:clusters} indicates, these clusters may co-exist (e.g. all three types co-exist in $A2$ alloy aged for 5 min). Also, $C2$ and $C3$ clusters are observed to co-exist in $A2$ alloy aged for 20 min (Fig.~\ref{fig:HRTEMcl}c). Moreover, $C3$ clusters are observed to co-exist with $GP$-$zone$ precipitates (Fig.~\ref{fig:HRTEMcl}f). These precipitates, which reveal various crystal symmetries relevant to pre-$\beta''$ phase reported by Marioara {\it et al}~\cite{marioara01,marioara03}, will be analyzed in a future report. However, the aging time scales and the nature of co-existence of these clusters and $GP$-$zones$ may suggest that the early-stage phenomena follow a sequential pattern, with $C1$ evolving to $C2$, $C2$ to $C3$ and $C3$ to $GP$-$zones$.  

\begin{table*}[ht]
\caption{Characteristics of various types of clusters identified through HRTEM-STEM observations.} 
\resizebox{6in}{!}{\includegraphics{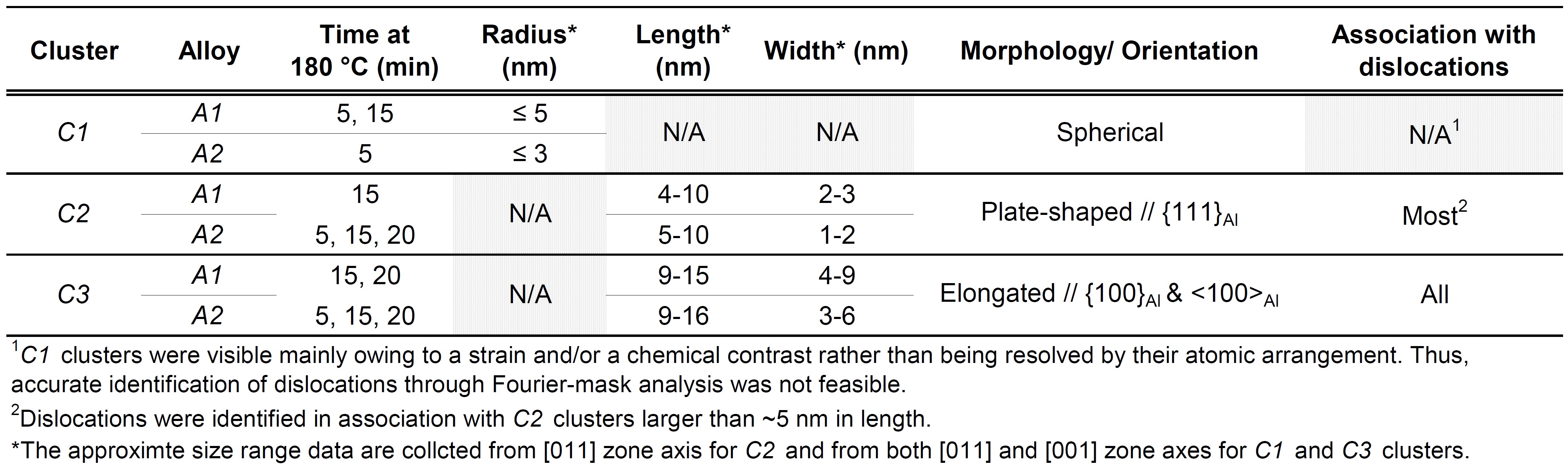}}
\label{table:clusters} 
\end{table*}    
  
\subsection{{\it PFC} simulation of early-stage clustering}   
To shed light on the evolutionary path of the clustering process, the early-stage clustering in an Al-Mg-Si supersaturated solid solution is further analyzed numerically via {\it PFC} simulations. Below, we first reproduce the thermodynamic properties of the system reconstructing the equilibrium phase diagram through adjustment of relevant {\it PFC} variables. It is then followed by numerical simulation of clustering guided through the estimated equilibrium properties of the system.     

\subsubsection{Phase diagram reconstruction}   
\begin{figure*}[htbp]
\resizebox{7.in}{!}{\includegraphics{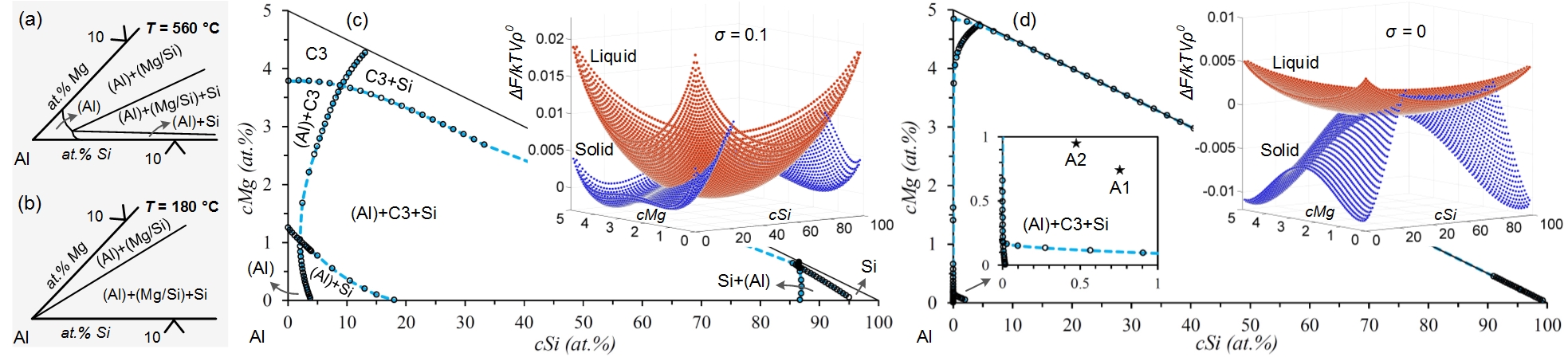}}
\caption{\textbf{Reproduction of thermodynamic properties.} (a-b) The Al-rich corner of an isothermal cut through the equilibrium Al-Mg-Si phase diagram at $T=560~^\circ$C and $T=180~^\circ$C, respectively, constructed using ThermoCalc software and database COST-507; Note: (Mg/Si) phase denotes all Mg-rich or Mg-Si rich phases present in the Al-Mg-Si system; (c-d) The phase diagram of the $(Al)$-$Si$-$C3$ system constructed using {\it PFC} free energy at model's reduced temperatures $\sigma=0.1$ and $\sigma=0$, respectively. The parameters for the ideal free energy and entropy of mixing are set as $\eta=1.4$, $\chi=1$ and $\omega=0.005$. The width of correlation peaks for all phases are taken as $\alpha_{111}=0.8$ and $\alpha_{100}=\frac{\sqrt[]{3}}{2}\alpha_{111}$. The chemical energy coefficient $\lambda$ in the interpolation functions (Eq.~\ref{interp-func}) is set to 0.03. For simplicity, the effective transition temperature for all families of planes in all phases, $\sigma_Mj$, is set to 0.55. The peak positions for various FCC phases in this study are set to $k_{111Al}=2\pi\sqrt{3}$, $k_{100Al}=\frac{2}{\sqrt{3}}k_{111Al}$, $k_{111C3}=(143/74)\pi\sqrt{3}$, $k_{100C3}=\frac{2}{\sqrt{3}}k_{111C3}$, $k_{111Si}=(572/271)\pi\sqrt{3}$ and $k_{100Si}=\frac{2}{\sqrt{3}}k_{111Si}$. The concentration $c_{Mg}$ in (c-d) is rescaled considering the maximum Mg-content of 5 $at.\%$ in $C3$.}
\label{fig:PhaseDiagram}
\end{figure*}

The numerical analysis begins with emulating the thermodynamic properties of the Al-rich portion of Al-Mg-Si system. We select the polynomial parameters in Eq.~(\ref{simplified-Energy}) (namely $\eta$, $\chi$ and $\omega$) and the width of various peaks ($\alpha_j$) in the correlation kernel $\hat{C}^{ii}_{2j}$ so as to approximately reproduce the equilibrium solubility limits of the solid solution $(Al)$ phase at solutionizing and aging temperatures (similar to the experimental phase diagrams shown in Figs.~\ref{fig:PhaseDiagram}a and~\ref{fig:PhaseDiagram}b, respectively). To introduce isotropic elastic constants in the solid phases, the ratio of the width of the correlation peaks for the two main families of planes in the fcc system are chosen $\alpha_{(111)}/\alpha_{(100)}=2/\sqrt{3}$~\cite{greenwood11}). The parameters used are given in the caption of Fig.~\ref{fig:PhaseDiagram}. We then select the target equilibrium properties that drive the kinetics of the PFC model in simulations. In our simulations, the last cluster phase with FCC structure to evolve, $C3$, is chosen to represent the final Mg-rich phase. The composition of this phase is approximated by a maximum of 5 $at.\%$Mg content to allow a reasonable numerical efficiency for clustering simulations in alloys with low average Mg contents. This approximation is supported by the notion that these clusters can be precursors to and have a lower Mg content than the GP-zones reported by Buha {\it et al.}~\cite{buha08-2} with an average 13.7 $at.\%$Mg content after a relatively long multi-step aging of Al-1.15Mg-0.6Si-0.12Cu ($at.\%$) alloy. The Si content of the $C3$ phase, depending on the selected average alloy composition of the phase diagram, can vary between zero and 5 $at.\%$ maximum, giving a possible minimum Mg/Si ratio of 1. Also, for simplifying the simulation, the Si-rich phase in the phase diagrams shown in Fig.~\ref{fig:PhaseDiagram} is assumed to be pure Si with an FCC structure rather than the known diamond cubic structure. Although this simplification influences the time scale for phase formation (i.e. shortens the computation time), it does not alter the evolutionary sequence of phases and Mg-rich nature of the second phase being evolved through the clustering transformation. This is mainly due to the chosen FCC lattice parameter of the $Si$ phase in {\it PFC} simulations in a way to introduce an infinitesimally small solubility of the Al matrix for Si at the aging temperature (i.e. matching the equilibrium phase diagram shown in Fig.~\ref{fig:PhaseDiagram}b). The composition-dependent lattice parameter of the solid phase is calculated below using the interpolation functions in Eq.~\ref{interp-func} interpolating between those of the three destination phases: $a_{(Al)}$, $a_{Si}$ and $a_{C3}$.  
\begin{align}
a_s = X_1a_{(Al)} + X_2a_{C3} + X_3a_{Si}
\label{latticeP}
\end{align}

The insets in Fig.~\ref{fig:PhaseDiagram}(c-d) show the calculated free energy landscapes of solid and liquid for temperature parameters of the model, i.e., the model's reduced temperatures $\sigma=0.1$ and $\sigma=0$ (which represent the solutionizing and aging temperatures of $560~^\circ C$ and $180~^\circ C$, respectively). Fig.~\ref{fig:PhaseDiagram}(c-d) illustrates the constructed isothermal sections of the ternary system comprising of $(Al)$, an Si-rich phase, $Si$, and the final cluster phase, $C3$. To construct the isothermal phase diagrams, the coexistence (solidus) lines for $(Al)-C3$, $(Al)-Si$, $C3-Si$ and $(Al)-C3-Si$ were obtained through satisfying equal chemical potential and grand potential for each species in the chosen phases. For example, the following set of equations were solved to obtain the $(Al)-C3$ coexistence line: 
\begin{align}
&\mu_{c_{Si}}^{(Al)}=\mu_{c_{Si}}^{C3}\nline&\mu_{c_{Mg}}^{(Al)}=\mu_{c_{Mg}}^{C3}
\nline& f^{(Al)}-\mu_{c_{Si}}^{(Al)}c_{Si}^{(Al)}-\mu_{c_{Mg}}^{(Al)}c_{Mg}^{(Al)}=f^{C3}-\mu_{c_{Si}}^{C3}c_{Si}^{C3}-\mu_{c_{Mg}}^{C3}c_{Mg}^{C3},
\label{coex}
\end{align}  
where $\mu_{c_{Si}}=\partial f/\partial {(c_{Si})}$ and $\mu_{c_{Mg}}=\partial f/\partial {(c_{Mg})}$ are the chemical potentials of the concentrations $c_{Si}$ and $c_{Mg}$, respectively. As Fig.~\ref{fig:PhaseDiagram}c shows, at the solutionizing temperature, both $A1$ and $A2$ alloy compositions fall in the single-phase region of $(Al)$. However, as demonstrated in Fig.~\ref{fig:PhaseDiagram}d, they lie within the multi-phase region of $(Al)+C3+Si$ at the aging temperature. Thus, at the aging temperature, both alloys with a non-equilibrium single-phase structure (i.e. upon quenching) must ideally seek equilibrium by decomposing into a multi-phase structure. 

\subsubsection{Numerical simulation}   
\begin{figure*}[htbp]
\resizebox{7.in}{!}{\includegraphics{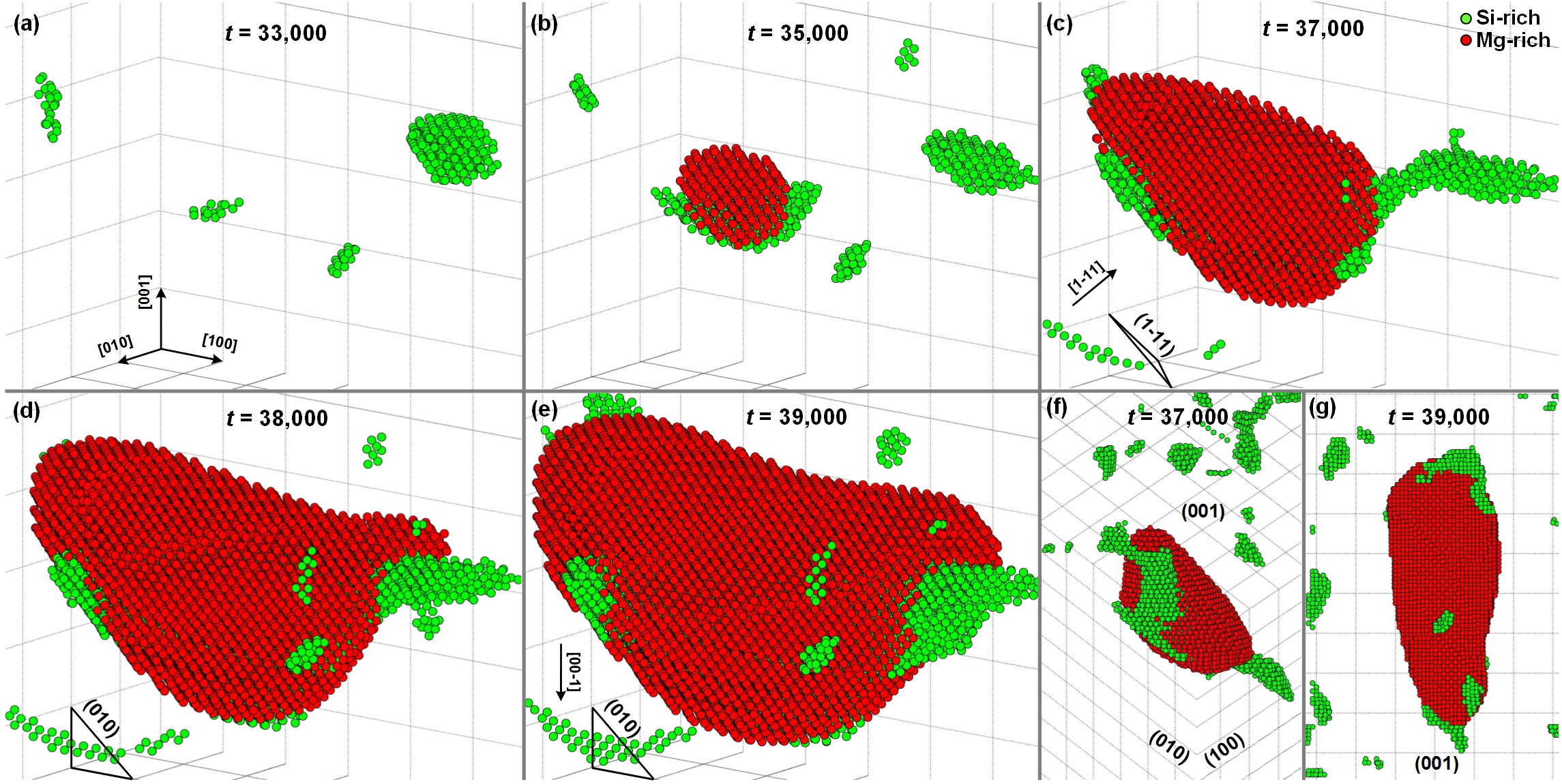}}
\caption{\textbf{Simulated formation of clusters.} (a-e) Early stage formation of typical surviving clusters during the {\it PFC} simulation of clustering in the $A2$ alloy; (f) The cluster shown in (c) viewed along [1$\bar{1}$1] axis; (g) The cluster shown in (e) viewed along [00$\bar{1}$] axis; To display the Mg- and Si-rich clusters, density peaks were filtered with minimum content of 2.5$at.\%$Mg and 4$at.\%$Si, respectively; These values represent approximately 50$\%$ and 100$\%$ of the maximum Mg and Si contents, respectively, in the $C3$ phase of the phase diagram shown in Fig.~\ref{fig:PhaseDiagram}d.}
\label{fig:PFCcl}
\end{figure*}

The simulation of clustering is carried out applying dissipative dynamics to the {\it PFC} free energy functions while introducing quenched-in dislocations~\cite{fallah13-2,fallah13,fallah12} into a single-phase $(Al)$ domain of $A1$ and $A2$ alloy compositions held at $\sigma=0$. A low amplitude stochastic noise is introduced to both density and concentration fields to simulate stochastic fluctuations that can potentially be required to stimulate nucleation. The 3D simulations were performed on a domain of 224$\times$224$\times$64 FCC lattice spacings, each resolved with 8 mesh points. The gradient energy coefficients $\alpha_{A}$ and $\alpha_{B}$ in Eq.~\ref{simplified-Energy} were both set to 0.6. Randomly-oriented partial edge dislocations (15 atoms in length) were introduced to the simulation domain at the quench temperature with a number density of $\approxeq$1 per 400 atoms. Fig.~\ref{fig:PFCcl} shows the simulation results within a small box taken from the simulation domain at various times during early-stage clustering. The nucleation and growth of few early clusters are captured in this box. Although the evolution of composition in the PFC simulations is quantitative, the presentation of mixing of Mg and Si atoms in Fig.~\ref{fig:PFCcl} is rather qualitative in nature. This is because the solute species, represented by coloured spheres in this figure, are representative of richness in the specific solute defined by the filtering criteria for the density peaks (which represent atoms in the PFC methodology). Based on such criteria, to display the Mg- and Si-rich clusters, density peaks have been filtered with minimum content of 2.5$at.\%$Mg and 4$at.\%$Si, respectively. With this qualitative presentation of mixing, Fig.~\ref{fig:PFCcl}(b) through Fig.~\ref{fig:PFCcl}(e) show that Mg atoms start to gather around an initially Si-rich cluster and gradually turn it into a Mg-rich cluster. It can also be seen in Fig.~\ref{fig:PFCcl}(e) that this cluster still shows colonies with a residual Si-rich nature. 

Simulation results reveal that, first, Si-rich clusters appear in the supersatureated matrix (Fig.~\ref{fig:PFCcl}a). Some of these clusters then attract a higher proportion of Mg and become Mg-rich clusters (Fig.~\ref{fig:PFCcl}b) with compositions close to that of the matrix (i.e. clusters are comprised of a high Al content). These Mg-rich clusters (shown in Fig.~\ref{fig:PFCcl}c-e) continue to grow in size and become increasingly enriched in Mg, while new Si-rich clusters can continue to nucleate in the matrix. Mg-rich clusters finally approach the assigned composition for $C3$ as specified for each alloy by the phase diagram in Fig.~\ref{fig:PhaseDiagram}d. As the Mg-rich (i.e. surviving) clusters grow in size, they assume preferred growth orientations and associated morphology changes, as well. The evolution path for a typical growing cluster in $A2$ alloy is demonstrated in Fig.~\ref{fig:PFCcl}. As Fig.~\ref{fig:PFCcl}b shows such a cluster appears with an initial spherical morphology. However, its further growth occurs preferentially first on the closed packed $\{111\}_{Al}$ planes (Fig.~\ref{fig:PFCcl}c), gradually evolving towards $\{100\}_{Al}$ $<$$100$$>_{Al}$ preferred orientation (Fig.~\ref{fig:PFCcl}d-e). The above evolution of size, orientation and morphology during formation of early clusters is in exact accordance with the corresponding characteristics of $C1$, $C2$ and $C3$ clusters. In addition, the simulation results show that $C1$ clusters may be both Si-rich and Mg-rich in nature. It should be noted that the very small size of the Si-rich clusters might prevent them from being detected by any electron microscopy examination.    

\section{Mechanisms of formation}
An analysis of cluster energy is undertaken within the framework of a {\it PFC} mean-field approximation of free energy to understand the mechanisms controlling the formation and evolution of the above identified clusters. 

\subsection{Work of formation}
The work of formation of clusters, $W_h$, is estimated by calculating its contributing terms, namely the chemical driving force, $\Delta f$, strain energy, $\Delta G_s$, and surface energy, $\gamma$~\cite{fallah12,fallah13,fallah13-2}.  
\begin{align}
W_h &= 2\pi R\gamma + \pi R^2 (-\Delta f + \Delta G_s)
\label{Hom.Nucl.En}
\end{align}
where $R$ is the cluster radius in terms of number of lattice spacings. The above quantities are estimated below for clustering in supersaturated Al-Mg-Si alloys of $A1$ and $A2$ compositions (denoted on the inset of Fig.~\ref{fig:PhaseDiagram}(d)). The evolving clusters are then either Mg- or Si-rich in nature. According to the simulation results in Fig.~\ref{fig:PFCcl}, the final clusters are rich in both Mg and Si representing a phase similar in composition to the $C3$ phase. We thus consider the quasi-equilibrium composition of $C3$ as the composition of the final cluster phase in the following calculations. The equilibrium compositions of $C3$ (i.e. $c_{Mg}^{cl}$ and $c_{Si}^{cl}$) and the corresponding bulk matrix (i.e. $c_{Mg}^{b}$ and $c_{Si}^{b}$) for each alloy composition on the phase diagram (i.e. $A1$ or $A2$ on Fig.~\ref{fig:PhaseDiagram}(d)) are determined by examining Eq.~\ref{coex} while also requiring mass conservation. Following Fallah {\it et al.}~\cite{fallah13}, the bulk driving force for the formation of $C3$ clusters with the equilibrium concentration is defined as 
\begin{align}
-\Delta f = f^b - \mu_{c_{Mg}}^b|_{c_{Mg}^b}(c_{Mg}^b-c_{Mg}^{cl})\nline -\mu_{c_{Si}}^b|_{c_{Si}^b}(c_{Si}^b-c_{Si}^{cl}) -f^{cl},
\label{DrivingForce}
\end{align}
where superscripts `$b$' and `$cl$' denote the bulk matrix and cluster ``phase'' quantities, respectively.

The strain energy associated with a coherent nucleus is calculated by~\cite{hoyt}:
\begin{align}
\Delta G_s = 2 G_A \delta^2 \frac{K_B}{K_B+G_A},
\label{StrainEnergy}
\end{align}
where 
\begin{align}
\delta = \frac{a_s^b-a_s^{cl}}{a_s^b}
\end{align}
is the misfit strain. $G_A$ and $K_B$ are 3D shear and bulk moduli, respectively, calculated from PFC mode approximation~\cite{greenwood11}. The free energy of an FCC crystal, in the limit of small deformations~\cite{elder10}, is evaluated at different strained states by substituting their respective coordinate transformations into a two-mode approximation of the density field and integrating over the corresponding strained unit cell. The elastic constants $C_{11}$, $C_{12}$ and $C_{44}$ ($C_{12}=C_{44}=C_{11}/3$) are then calculated through fitting the resultant free energy to parabolic expansions in displacement fields~\cite{greenwood11,fallah12,fallah13}. The 3D shear and bulk moduli can be simply calculated through solving the following set of equations~\cite{greenwood11,fallah13-2},
\begin{align}
&G_A = C_{44}\nline
&\nu = \frac{C_{12}}{C_{11}+C_{12}}\nline
&E = 2G_A(1+\nu)\nline
&K_B = \frac{E}{3(1-2\nu)},
\label{ElConst}
\end{align}
where $E$ is the Young's modulus and $\nu$ is the Poisson ratio. Following the approach presented in Ref.\cite{hoyt}, the bulk modulus is calculated for the cluster composition, whereas the shear modulus is evaluated for the bulk matrix composition.    

Neglecting the structural contributions, the interfacial free energy of a coherent cluster is taken to be solely chemical~\cite{Turnbull}, assuming a low dislocation density in the system. Following Cahn and Hilliard~\cite{cahn58}, the interfacial free energy between a $C3$ cluster and the Al-rich matrix is evaluated by the following analytical form:   
\begin{align}
\gamma &= 2\int\limits_{c_{Mg}^b}^{c_{Mg}^{cl}}{\bigg[\alpha_{Mg} (f-f^b)\bigg]^\frac{1}{2}\bigg\{1+(\frac{\alpha_{Si}}{\alpha_{Mg}})(\dxdy{c_{Si}}{c_{Mg}})^2\bigg\}^\frac{1}{2}}dc_{Mg},
\label{Surf.E}
\end{align}
where $\alpha_{Mg}$ and $\alpha_{Si}$ are gradient energy coefficients for Mg and Si, respectively, both set to 0.6 in this study. The term $\dxdy{c_{Si}}{c_{Mg}}$ is estimated by $-\frac{\frac{\partial f}{\partial c_{Mg}}}{\frac{\partial f}{\partial c_{Si}}}$ and the variation of $c_{Si}$ with respect to $c_{Mg}$ is approximated by a linear interpolation between the bulk and the equilibrium cluster compositions (i.e., $c_{Si}=c_{Si}^b+\frac{c_{Si}^{cl}-c_{Si}^b}{c_{Mg}^{cl}-c_{Mg}^b}(c_{Mg}-c_{Mg}^b)$).

\begin{figure}[htbp]
\resizebox{3.4in}{!}{\includegraphics{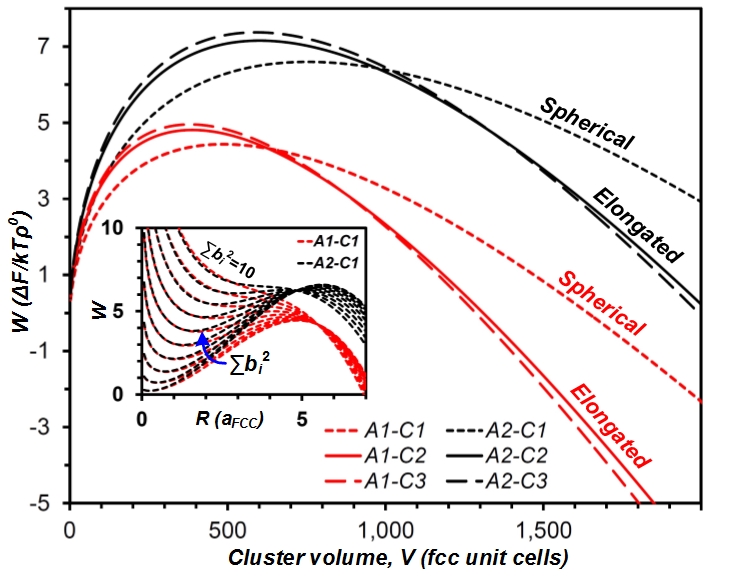}}
\caption{\textbf{Energy analysis of clustering.} The work of formation of a spherical ($C1$) and elongated clusters ($C2$ and $C3$) in $A1$ and $A2$ alloys; The inset displays the dislocation-assisted removal of the energy barrier for clustering in the case of spherical morphology.}
\label{fig:Wsph}
\end{figure}

To analyze the energetic mechanisms underlying the morphology and orientation change of early clusters, the work of formation is calculated for a sphere and a spheroid of $r_2/r_1=1.5$ lying on either $\{111\}_{Al}$ or $\{100\}_{Al}$ orientations. To be comparable to various observed morphologies, the estimated work of formation in Fig.~\ref{fig:Wsph} is plotted as a function of cluster volume (i.e. the total number of enclosed unit cells) by taking $R=(\frac{3V}{4\pi}.\frac{1}{C_{or}})^{-1/3}$. $C_{or}$ is a coefficient that takes into account the effect of orientation on the number of atoms that can fit into a given cluster volume. For a sphere, $C_{or}$ equals 1 while, using a geometrical model, it is estimated to be 1.052 and 1.034 for spheroids along $\{111\}_{Al}$ and $\{100\}_{Al}$ orientations, respectively. As Fig.~\ref{fig:Wsph} shows, initially, the spherical cluster assumes the lowest total free energy in both alloys. However, beyond a certain cluster volume, the cluster assumes an elongated (i.e. spheroidal) morphology in order to lower the total free energy. Based on the calculated $C_{or}$ values, the elongated morphology for a coherent cluster of a given volume has a smaller surface to volume ratio at larger cluster sizes where the contribution of surface energy to the total free energy of cluster becomes dominant. Furthermore, the impact of different orientations is embedded in the values of $C_{or}$ in such a way that an elongated, coherent cluster of a given volume possesses a larger surface energy while lying on the more compact planes of $\{111\}_{Al}$. The resultant calculations, as shown in Fig.~\ref{fig:Wsph}, shows that an orientation change from $\{111\}_{Al}$ to $\{100\}_{Al}$ with the growth of elongated morphology clusters results in a total free energy reduction. The larger size ranges of the experimentally observed $C3$ clusters and the absence of $C2$ clusters in $A1$ alloy after 20 min aging, as compared in Table~\ref{table:clusters}, also support this finding.  

\begin{figure*}[htbp]
\resizebox{7.in}{!}{\includegraphics{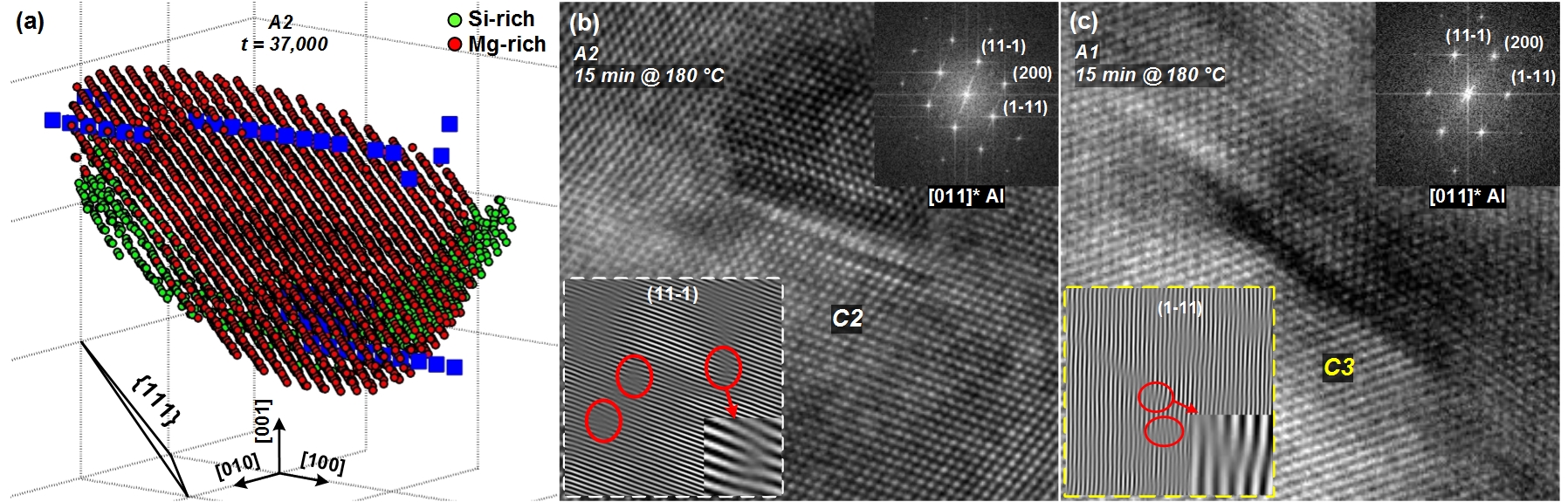}}
\caption{\textbf{Association of clusters with dislocations.} (a) a simulated cluster in $A2$ alloy at $t=37,000$, and (b-c) the observed $C2$ and $C3$ clusters in Fig.~\ref{fig:HRTEMcl}c-d, respectively; The inset Fourier-masked micrographs reveal dislocations marked with circles; In (a), the blue square symbols represent the locus of dislocation lines associated with the cluster; Also, to display the Mg- and Si-rich clusters, density peaks were filtered with minimum content of 2.5$at.\%$Mg and 4$at.\%$Si, respectively.}
\label{fig:ClStruc}
\end{figure*}

Furthermore, the simulation results have indicated that the $A1$ and $A2$ alloys, while exhibiting the same sequence of early clustering (i.e. $C1$, $C2$, $C3$), differ in terms of the kinetics of transformation, i.e. in the evolution of number density. This is supported by the clustering energy analysis (plotted for the two alloys in Fig.~\ref{fig:Wsph}), indicating that the alloy with smaller Mg/Si ratio (i.e. the Si-rich alloy, $A1$) exhibits a smaller energy barrier and also a smaller critical size for nucleation of clusters. This leads to a larger number density of clusters in the Si-rich alloy agreeing with the APT results of Torsæter {\it et al}~\cite{torsaeter10}. Our simulation results also indicate that the clusters continuously evolve in chemical composition (as Fig.~\ref{fig:PFCcl} shows) with the advancing of the clustering process. The evolution of chemical composition and number density of clusters in various alloys is the subject of our upcoming study including {\it PFC} simulations verified against new APT results.

\subsection{Effect of dislocations}
The following extended form of the work of formation has been shown to incorporate the effect of dislocations on the formation of clusters~\cite{fallah12,fallah13,fallah13-2}: 
\begin{align}
W_d &= W_h - \Delta G_{sr} + \Delta G_d \nline&= 2\pi R\gamma + \pi R^2 (-\Delta f + \Delta G_s) - \Delta G_{sr} + \Delta G_d.
\label{Dis.Nucl.En}
\end{align}
$\Delta G_{sr}$ is the stress relaxation term due to segregation of solute atoms into dislocations~\cite{cahn57}, which is described here by
\begin{align}
\Delta G_{sr} = \psi^2\chi_d E A \ln(R),
\label{StressRelaxE}
\end{align}

where $A= \frac{G_A \Sigma b_i^2}{4 \pi (1-\nu)}$, $\psi$ is the linear expansion coefficient with respect to composition, $\chi_d$ represents the change in the diffusion potentials as a function of composition. We introduce $\Sigma b_i^2$ to represent a weighted average of the magnitude of Burger's vectors around the dislocations accompanying the cluster. The prefactor $\psi^2\chi_d E A$, which accounts for the strain energy reduction due to solute segregation around a dislocation~\cite{larche85}, is approximated by~\cite{fallah13} 
\begin{align}
\psi^2 \chi_d = \frac{\psi_{Mg} \psi_{Mg} \frac{\partial^2 f}{\partial c_{Si}^2} + \psi_{Si}\psi_{Si} \frac{\partial^2 f}{\partial c_{Mg}^2} +2 \psi_{Si} \psi_{Mg} \frac{\partial^2 f}{\partial c_{Si} \partial c_{Mg}}}{\frac{\partial^2 f}{\partial c_{Si}^2}\frac{\partial^2 f}{\partial c_{Mg}^2}-\left(\frac{\partial^2 f}{\partial c_{Si} \partial c_{Mg} }\right)^2},
\label{etachi}
\end{align}
where $\psi_{Si}=-\frac{1}{a}\frac{\partial a}{\partial c_{Si}}$ and $\psi_{Mg}=-\frac{1}{a}\frac{\partial a}{\partial c_{Mg}}$. 

The term $\Delta G_d$ in Eq.(\ref{Dis.Nucl.En}) takes into account the increase in the total energy of the system due to presence of dislocations:
\begin{align}
\Delta G_d = \zeta A,
\label{Disl.E}
\end{align}
where $\zeta$ is a prefactor of the order 10, representing the average amount of energy per dislocation core~\cite{Hull}.

The analysis of Fallah {\it et al}~\cite{fallah12,fallah13,fallah13-2} on Al-Cu-(Mg) system clearly showed that the presence of quenched-in dislocations in the matrix of a supersaturated solid solution system could lead to stress relaxation effects associated with the sum of dislocation Burger's vectors, $\Sigma b_i^2$, and thus effectively reduce, or even eliminate, the local energy barrier for nucleation. This effect can be similarly shown in the current system (see the inset of Fig.~\ref{fig:Wsph}). Fig.~\ref{fig:ClStruc}a shows a typical simulated spheroidal cluster in direct association with dislocations in the surrounding lattice. To confirm the validity of such cluster-dislocation associations in the Al-Mg-Si system, HRTEM images are further analyzed by Fourier masking method. As demonstrated in Fig.~\ref{fig:ClStruc}b-c, and indicated in Table~\ref{table:clusters}, the analysis confirms that most clusters are indeed surrounded by dislocations. On larger scales, HRTEM images, such as Fig.~\ref{fig:HRTEMcl}c-d, show evidence for the formation of clusters along high strain paths which resemble such paths for dislocation arrangements~\cite{pogatscher13}. From the above results, we propose that the analysis of the earliest-stage direct clustering, and thus the kinetics of precipitation, in Al-Mg-Si alloys must take into account the presence of dislocations and their potential interactions with clustering nucleation processes. The current observations, however, do not rule out the possibility of other mechanisms being active during clustering. It has been long believed that quenched-in vacancies, which exist in large fractions upon quenching, play an important role in clustering. Cluster nucleation or evolution with no defects involved may not also be ruled out here (it can be possible if the high energy barrier as shown in Fig.~\ref{fig:Wsph} is overcome). The important conclusion here is that dislocations have been directly observed and identified in association with the clusters. 

The findings of this work will be useful in future bottom-up design of desirable structures and cost-effective development of alloys, processes and properties.\\       

\section{Summary}
In this work, detailed high resolution electron microscopy and a highly-versatile 3D computational modelling, i.e. {\it PFC}, approach have been used to illuminate the profoundly ambiguous nature of early-stage precipitation in AA6000 Aluminium alloys. Through the microscopy analysis, it is discovered that the earliest clusters, all with FCC structure, appear with spherical morphology ($C1$) and then evolve into elongated morphology, first, lying on the close-packed $\{111\}_{Al}$ ($C2$), and then, orienting on the $\{100\}_{Al}$ planes and along the $<$$100$$>_{Al}$ directions ($C3$). These findings are in full agreement with the simulation results that clearly identifies the sequential nature of clustering processes as: $C1$ $\rightarrow C2 // \{111\}_{Al} \rightarrow C3 // \{100\}_{Al}$ $<$$100$$>_{Al}$. The experimental observations also suggest that $C3$ clusters evolve into similarly-oriented $GP$-$zone$ precipitates with a crystal structure deviating from FCC. Further work of formation analysis using {\it PFC} simulations explains the energetic reasons behind the cluster evolution. This work asserts that quenched-in dislocations are critical for mediating the cluster nucleation and growth processes.

The systematic knowledge developed here opens a new gateway through which desirable strengthening can effectively be achieved through controlling early-stage clustering phenomena. This is of significant practicality in addressing the intense desire for designing new generations of light-weight automotive components and cost-environment dictated short-time, low-temperatures paint baking processes.              

\begin{acknowledgements}
We acknowledge the financial support received from Ontario Ministry of Research and Innovation (Early Researcher Award Program), Novelis Inc., Initiative for Automotive Manufacturing Innovation (IAMI), National Science and Engineering Research Council of Canada (NSERC) and the Clumeq High Performance Centre. Electron microscopy was performed at the Canadian Centre for Electron Microscopy (also supported by NSERC and other government agencies). We also thank Dr. Leo Colley for his contribution to alloy design and fabrication and Dr. Brian Langelier for helpful insight on electron microscopy.
\end{acknowledgements}


\bibliography{references}

\end{document}